\documentclass[10.7pt,aps,twocolumn,nofootinbib]{revtex4}
\usepackage{amsmath,graphics,epsfig}
\usepackage{color,hyperref}
\usepackage{datetime}
\begin{document}

\title{Extended MSSM in Supersymmetric $\rm{SO}(10)$ Grand Unification}
\author{Sibo Zheng}
\email{sibozheng.zju@gmail.com}
\affiliation{Department of Physics, Chongqing University, Chongqing 401331, P. R. China}
\date{August 3, 2018}

\begin{abstract}
We apply the perturbative grand unification due to renormalization to distinguish TeV-scale relics of supersymmetric $\rm{SO}(10)$ scenarios. 
With rational theoretical constraints taken into account, 
we find that for the breaking pattern of either $\rm{SU}(5)$ or Pati-Salam 
only extra matter $\mathbf{16}$ supermultiplet of $SO(10)$ can appear at TeV scale, apart from MSSM spectrum.
\end{abstract}
\maketitle

\section{Introduction}
The discovery of standard model (SM)-like Higgs \cite{1207.7214,1207.7235} 
provides a new portal to TeV-scale new physics at the LHC in the forthcoming  years.
Among other things, such new physics models may reveal the ``nature" of SM-like Higgs,
and offer a novel mechanism to stabilize divergence involving SM Higgs.
For those interesting scenarios in the literature, in this paper we are restricted to the idea of supersymmetry (SUSY).
Specifically, we will utilize the grand unification (GUT) \cite{GG}, 
which is one of the most beautiful features delivered by SUSY,
to distinguish TeV-scale relics of SUSY GUT models.
For reviews on this subject see e.g. \cite{Langacker, SlanskyYR}.

In the viewpoint of unification, 
the minimal supersymmetric standard model (MSSM) can be embedded into conventional $SU(5)$ \cite{DG,Sakai,DRW},  
$SO(10)$ \cite{Georgi,FM} or other GUT models with gauge groups of higher ranks.
In the light of our previous study on $SU(5)$ \cite{1706.01071},
we will continue to explore the TeV-scale relics of SUSY $SO(10)$ unification. 
Comparing with $SU(5)$, 
the low-energy effective theories of $SO(10)$ are more complex.
The first major reason is that there may be multiple intermediate scales between the weak and GUT scale.
The second reason is that
since a lot of higher-dimensional representations of $SO(10)$ trivially satisfy gauge anomaly free condition,
the constraint imposed by  this condition is much weaker in $SO(10)$. 
Earlier studies on low-energy effective theory which is consistent with perturbative SUSY $SO(10)$ unification 
are based on specific motivations such as Higgs mass \cite{0807.3055} and neutrino physics \cite{9209215, 0408139, 0910.3924}.

Instead of particular phenomenological concerns,
we will take a systematic analysis on the low-energy effective theory.
In order to simplify the analysis on extra matter beyond the MSSM spectrum, 
we will explore $SO(10)$ scenarios with the following theoretical features.
\begin{itemize}
\item The $SO(10)$ unification is strictly perturbative.
 \item In the chain of gauge symmetry breaking
 \begin{eqnarray}{\label{scenario}}
 SO(10) \xrightarrow{H_{1}} G_{1} \xrightarrow{H_{2}} G_{2}\cdots \xrightarrow{H_{n}} G_{\text{SM}},
\end{eqnarray}
where $G_{\text{SM}}$ refers to the SM gauge group.
When the Higgs component fields responsible for two nearby steps of gauge symmetry breaking 
can be contained in a single Higgs supermultiplet,
these two Higgs supermultiplets will be identified as the same one.
Otherwise, they differ from each other 
\footnote{In this situation, the splitting between two nearby scales of gauge symmetry breaking is generally large.
We will impose the relatively strong constraint that the two Higgs supermultiplets responsible for such large splitting 
are forbidden to directly couple to each other.}.
 \item In order to avoid dangerous mixings among Higgs vevs $\left<H_{i}\right>$,
 all of $H_i$ are forbidden to directly couple to each other.
 \item In order to avoid dangerous masses or mixing effects, neither the MSSM fields nor extra matters are allowed to directly 
couple to any Higgs supermultiplets $H_{i}$ in Eq. (\ref{scenario}). 
\footnote{There is a loophole. When the vev of the last Higgs $\left<H_{n}\right>$ is of order TeV scale,
they are allowed to couple to $H_{n}$.}
 \end{itemize}
The main reason for the last point is that the vacuum expectations (vevs) of $\left<H_{i}\right>$ would 
result in large matter masses or large mixing effects if they were directly coupled to either MSSM fields or extra matters,
which would lead to them playing no role at weak scale.
For example, coupling a $\mathbf{54}$, which can break $SU(5)$ to SM gauge group, 
 to MSSM Higgs $\mathbf{10_{H}}$ through interaction $\mathbf{54}\times\mathbf{10_{H}}\times\mathbf{10_{H}}$,
yields unfavorable Higgsino mass for a large vev of SM singlet in $\left<\mathbf{54}\right>$.

Theoretical constraints above have been partially imposed in the literature to our knowledge.
However, they have never been combined together to derive a systematic analysis on the low-energy effective theory.
The paper is organized as follows.
In Sec.II,  we discuss the extra matter supermultiplets which are consistent with our starting points
in two well known patterns of gauge symmetry breaking.
In Sec.III,  we examine  the perturbative unification with these representations. 
Finally, we conclude in Sec.IV.

\section{Representations}
According to our starting points, in this section 
we investigate the representation of extra matter which can directly couple to SM Higgs $\mathbf{10_{H}}$ 
in the following two patterns of gauge symmetry breaking,
\begin{eqnarray}{\label{pattern}}
\rm{A} : SO(10) &\xrightarrow {H_{1}} & SU(5)  \xrightarrow{H_{2}} G_{\text{SM}},\nonumber\\
\rm{B} : SO(10) & \xrightarrow{H_{1}}  & SU(4)_{c}\times SU(2)_L \times SU(2)_R\nonumber\\
&\xrightarrow{H_{2}}& SU(3)_{c}\times SU(2)_L \times SU(2)_R \times U(1)_{B-L} \nonumber\\
&\xrightarrow{H_{3}}&  G_{\text{SM}}.\nonumber
\end{eqnarray}
Pattern A \cite{GN1,GN2, GN3, Rajpoot} is a two-step breaking with $SU(5)$ subgroup,
and pattern B \cite{ CEG, PS,M} is a three-step breaking referred to Pati-Salam model \cite{PS}.

Note, in the MSSM the SM fermion matters are described by $\mathbf{16}_{i}$ of $SO(10)$ with index $i=1-3$,
and the SM Higgs is contained in the $\mathbf{10_{H}}$ of $SO(10)$.
In particular, $\mathbf{16}_{i}$  contain three-generation right-hand neutrinos,
whereas $\mathbf{10_{H}}$ is composed of $\mathbf{5_{H}}$ and $\mathbf{\bar{5}_{H}}$ of $SU(5)$
which contain the two Higgs doublets of MSSM and two color-triplets.

\subsection{$SU(5)$}
In this pattern of symmetry breaking $H_{1}$ should contain an $SU(5)$ singlet,
there are two candidates $H_{1}=\{\mathbf{16},\mathbf{126}\}$.
The second Higgs $H_{2}$ should contain a $\mathbf{24}$ of $SU(5)$,
which corresponds to three potential choices $H_{2}=\{\mathbf{45},\mathbf{54}, \mathbf{210}\}$.
Since $H_{1}\neq H_{2}$, we take the rational that the splitting between these two broken scales is large.

With potential assignments on $H_{1}$ and $H_2$ above, there are six sets of combinations.
In each case, there may exist four types of dangerous gauge-invariant superpotentials 
which violate the last two starting points in the Sec.I: 
\begin{eqnarray}{\label{unsafe1}}
H_{1} &\times&\mathcal{N}\times \mathcal{N}; \nonumber\\
H_{2} &\times& \mathcal{N}\times \mathcal{N}; \nonumber\\
H_{1} &\times& H_{2} \times \mathcal{N}; \nonumber\\
H_{1} &\times& H_{1} \times  H_{2},  ~~~H_{2} \times H_{2} \times H_{1};
\end{eqnarray}
where  MSSM matter field $\mathcal{N}=\{\mathbf{16}_{i},\mathbf{10_{H}}\}$.
In Eq.(\ref{unsafe1}), the first three types of superpotentials tender to seed Dirac or Majorana masses to MSSM matters or MSSM Higgs doublets;
and the last type of gauge invariant superpotentials yields dangerous mixings between vevs of $H_{1}$ and $H_2$.

Firstly, one finds a dangerous operator $\mathbf{16}\times\mathbf{16}_{i}\times\mathbf{10_{H}}$ of the first type in Eq.(\ref{unsafe1})
which disfavors the choice $H_{1}=\mathbf{16}$.
Secondly, due to dangerous operator $\mathbf{126}\times\mathbf{210}\times\mathbf{10_{H}}$ the set of $H_{1}=\mathbf{126}$ and $H_{2}=\mathbf{210}$ is also disfavored.
Therefore, we are left with two combinations $H_{1}=\mathbf{126}$ and $H_{2}=\{\mathbf{45}, \mathbf{54}\}$.
In the case $(H_{1},H_{2})=(\mathbf{126},\mathbf{54})$, 
there exists a dangerous operator $\mathbf{126}\times\mathbf{126}\times\mathbf{54}$.
In the last case $(H_{1},H_{2})=(\mathbf{126},\mathbf{45})$, 
denote the new matter supermultiplets with $\mathcal{M}$,
we find that unsafe superpotentials exclude $\mathcal{M}=\{\mathbf{10}, \bar{\mathbf{16}}, \mathbf{120},\mathbf{144},\bar{\mathbf{144}}, \mathbf{210}\}$, 
leaving us only two possibilities, 
\begin{eqnarray}{\label{s1}}
\mathbf{16}_{M}\times\mathbf{16}_{M}\times\mathbf{10_{H}},
 ~~~~~~\mathbf{16}_{M}\times\mathbf{16}_{i}\times\mathbf{10_{H}}.
\end{eqnarray}

In compared with breaking pattern A,
there is another pattern of two-step breaking \footnote{The author thanks the referee for reminding us this case.} 
\begin{eqnarray}
SO(10)\xrightarrow {H_{1}} G_{\text{SM}}\times U(1)\xrightarrow{H_{2}}G_{\text{SM}}\nonumber.
\end{eqnarray}
In this case, the potential choices are $H_{1}=\{\mathbf{45},\mathbf{210}\}$ and $H_{2}=\{\mathbf{16},\mathbf{126}\}$.
According to Eq.(\ref{unsafe1}), 
dangerous operator $\mathbf{16}\times \mathbf{16}_{i}\times\mathbf{10_{H}}$ excludes the case $H_{2}=\mathbf{16}$.
Morevover, a dangerous operator $\mathbf{210}\times \mathbf{126}\times\mathbf{10_{H}}$ excludes $H_{1}=\mathbf{210}$.
Therefore, there is only a viable combination $(H_{1}, H_{2})=(\mathbf{45}, \mathbf{126})$,
in which case the extra matter $\mathcal{M}$ is similar to those of $SU(5)$ subgroup.

\subsection{Pati-Salam}
In this pattern of symmetry breaking, $H_{1}$ should contain a singlet of $SU(4)_{c}\times SU(2)_{L}\times SU(2)_{R}$,
which has two choices $H_{1}=\{\mathbf{54}, \mathbf{210}\}$.
$H_{2}$ should contain a singlet of $SU(3)_{c}$ and $U(1)_{B-L}$,
which is a $\mathbf{15}$ of $SU(4)_{c}$. 
There are two representations $H_{2}=\{\mathbf{45}, \mathbf{210}\}$ of $SO(10)$ which include such a $\mathbf{15}$.
Finally, $H_{3}=\mathbf{16}$ offers the breaking of $SU(2)_{R}\times U(1)_{B-L}\rightarrow U(1)_{Y}$.

Since $\mathbf{210}$ contains both a singlet of $SU(4)_{c}\times SU(2)_{L}\times SU(2)_{R}$ 
and a $\mathbf{15}$ of $SU(4)_{c}$, 
according to the second starting point both $H_{1}$ and $H_{2}$ are identified as $\mathbf{210}$.
In this case, the three-step breaking is approximately two-step.
With $H_1=H_2=\mathbf{210}$ we are left with a single choice $(H_{1}, H_{2}, H_{3})=(\mathbf{210}, \mathbf{210},\mathbf{16})$,
where similar to previous discussions about $SU(5)$ there are four viable choices for extra matters,
\begin{eqnarray}{\label{s2}}
\mathbf{16}_{M}&\times&\mathbf{16}_{M}\times\mathbf{10_{H}},~~~~
\mathbf{10}_{M}\times\mathbf{54}_{M}\times\mathbf{10_{H}} ,\nonumber\\
\mathbf{16}_{M}&\times&\mathbf{144}_{M}\times\mathbf{10_{H}},~~
\mathbf{144}_{M}\times\bar{\mathbf{144}}_{M}\times\mathbf{10_{H}}.
\end{eqnarray}
Note, unlike in pattern A, extra matter supermultiplets in Eq.(\ref{s2}) are allowed to directly couple to $H_{3}$.
Because the broken scale of $SU(2)_{R}\times U(1)_{B-L}$ \cite{PS, SU2m1,SU2m2} can be close to TeV scale (see, e.g. \cite{SU2c1,SU2c2,SU2c3,SU2c4,SU2c5,SU2c6}).

\section{Perturbative Unification}
With the theoretical constraints in the Introduction,
we have clarified that a single or two $\mathbf{16}$ supermultiplets are allowed in pattern A,
whereas two $\mathbf{16}$s,  a $\mathbf{10}$ with $\mathbf{54}$, a $\mathbf{16}$ with $\mathbf{144}$ or a pair of  
vector-like $\mathbf{144}$ may appear in the pattern B.
Now, we examine whether any of them are consistent with the first constraint - perturbative unification.

We start with the one-loop renormalization group equations (RGEs) for SM gauge coupling constant,
\begin{eqnarray}{\label{RGE}}
\frac{d}{dt}\alpha^{-1}_{i}=-\frac{b_{i}}{2\pi},
\end{eqnarray}
where RG scale $t=\text{ln}\mu$ and coefficients $b_{i}=(b_{U(1)_{Y}}, b_{SU(2)_{L}}, b_{SU(3)_{c}})$ 
are determined by \cite{Machacek1,Machacek2},
\begin{eqnarray}{\label{b}}
b_{i}=-\{\frac{11}{3}C^{i}_{2}(G)-\frac{4}{3}\cdot\kappa\cdot T(r_{f_{i}})-\frac{1}{6}T(r_{s_{i}})\}.
\end{eqnarray}
Here, $C_{2}(G)$ is the quadratic Casimir invariant, and $T(r)$ refers to dynkin index that depends on details of the representation \cite{SlanskyYR}.

\subsection{$SU(5)$}
\begin{figure}
\includegraphics[width=0.45\textwidth]{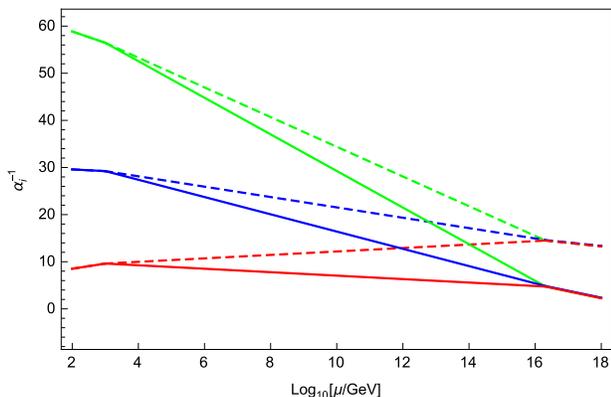}
\vspace{-0.45cm}
 \caption{One-loop RGEs for SM gauge coupling $\alpha^{-1}_{1}$ (green), 
$\alpha^{-1}_{2}$ (blue) and  $\alpha^{-1}_{3}$ (red) for extra matter $\mathbf{16}_{M}$ (dashed) 
and $\mathbf{16}_{M}+\mathbf{16}_{M}$ (solid), respectively. Here SUSY breaking scale $\Lambda_{\text{SUSY}}=1$ TeV.}
\label{su5}
\end{figure}

In the case of $SU(5)$ subgroup there are two intermediate scales $\Lambda_{\text{SUSY}}$ and $\Lambda_{5}$ 
between $M_{Z}$ and $\Lambda_{10}$, corresponding to SUSY and $SU(5)$  breaking, respectively.
The $b_i$ coefficients are given by \footnote{In order to eliminate the triplet fields in $\mathbf{10_{H}}$ some additional matter representations may be required as well.
Nevertheless, they are usually decoupled below the scale $\Lambda_{5}$, which don't affect the RG analysis between $M_{Z}$ and $\Lambda_{5}$.}
\begin{widetext}
\begin{eqnarray}{\label{b1}}
b_{i}&=&\begin{cases}
(41/10, -19/6, -7), & \mu\in [M_{\text{Z}}, \Lambda_{\text{SUSY}}] \\
(33/5+\delta b_{1}(\mathbf{M}), 1+\delta b_{2L}(\mathbf{M}), -3+\delta b_{3}(\mathbf{M})),   &    \mu\in [\Lambda_{\text{SUSY}},\Lambda_{5}] \\
 \end{cases}\nonumber\\
 b_{5}&=&~~\delta b_{5}(\mathbf{M}),
 ~~ ~~~~~~~~~~~~~~~~~~~~~~~~~~~~~~~~~~~~~~~~~~~~~~~~~~ \mu\in [\Lambda_{5}, \Lambda_{10}]
 \end{eqnarray}
 \end{widetext}
where $b_{5}$ denotes the $b$ coefficient of $SU(5)$ subgroup, 
and $\delta b_{i} (\mathbf{M})$ refer to contributions to $b$ coefficients due to extra matter.
In particular, $\delta b_{i}(\mathbf{16}_{M})=(2,2,2)$ and $\delta b_{i}(\mathbf{16}_{M}+\mathbf{16}_{M})=(4,4,4)$, respectively in Eq.(\ref{b1}).
Fig.\ref{su5} shows the plots of RG running of SM gauge coupling constants according to Eq.(\ref{b1}).
It reveals that for $\Lambda_{\text{SUSY}}=1$ TeV the $SU(5)$ unification occurs at $\Lambda_{5}\simeq 10^{16.3}$ GeV.
Moreover, the $SO(10)$ unification in both cases can occur at $\Lambda_{10}$ large than $10^{18}$ GeV.
Comparing $\Lambda_{5}$ with $\Lambda_{10}$, one finds that there is indeed sufficient splitting between them,
which verifies previous arguments.\\

\subsection{Pati-Salam}
In the case of Pati-Salam model there are two intermediate scales $\Lambda_{\text{SUSY}}$ and $\Lambda_{\text{R}}$ between $M_{Z}$ and $\Lambda_{10}$,  which denotes SUSY and $SU(2)_{R}\times U(1)_{B-L}$ breaking scale, respectively.
In this case the coefficients $b_{i}$ are given by,
\begin{widetext}
\begin{eqnarray}{\label{b21}}
b_{i}=\begin{cases}
(41/10, -19/6, -7), & \mu\in [M_{\text{Z}}, \Lambda_{\text{SUSY}}] \\
(33/5+\delta b_{1}(\mathbf{M}), 1+\delta b_{2}(\mathbf{M}), -3+\delta b_{3}(\mathbf{M})),   &    \mu\in [\Lambda_{\text{SUSY}},\Lambda_{\text{R}}] \\
 \end{cases}
 \end{eqnarray}
 \end{widetext}
for the RG scale between $M_{Z}$ and $\Lambda_{\text{R}}$, and 
\begin{widetext}
 \begin{eqnarray}{\label{b22}}
 b_{i}=(6+2+\delta b_{B-L}(\mathbf{M}), 0+2+\delta b_{2L}(\mathbf{M}),0+2+\delta b_{2R}(\mathbf{M}),
-3+2+\delta b_{3}(\mathbf{M})),~~ \mu\in [\Lambda_{R}, \Lambda_{10}] \nonumber\\
\end{eqnarray}
\end{widetext}
for the RG scale between $\Lambda_{\text{R}}$ and $\Lambda_{10}$.
Above the RG scale $\Lambda_{R}$, MSSM matters and Higgs field $H_{3}=\mathbf{16}$ contributes to 
$\delta b_{i}=(6,0,0,-3)$ and $\delta b_{i}=(2,2,2,2)$ in Eq.(\ref{b22}), respectively.
Regardless of what extra matter appears above $\Lambda_{\text{SUSY}}$
and what kind of Higgs $H_{1,2, 3}$ above $\Lambda_{\text{R}}$, 
$SO(10)$ unification at the one-loop level yields $\ln (\Lambda_{10}/\Lambda_{\text{R}})\simeq 2.53$ \cite{9602391,9911272},
which uniquely determines $\Lambda_{\text{R}}$ once the content of extra matters is identified.

\begin{figure}
\includegraphics[width=0.45\textwidth]{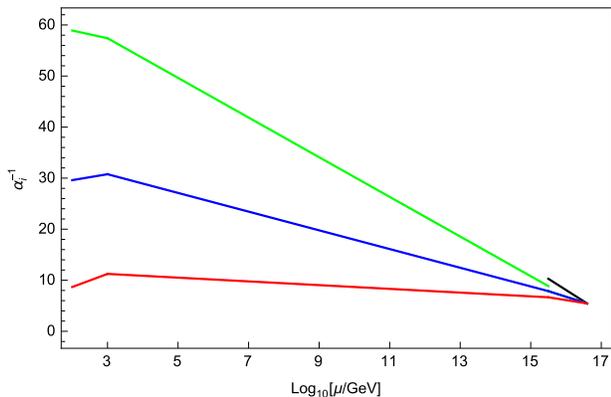}
\vspace{-0.45cm}
 \caption{One-loop RGEs for SM gauge coupling $\alpha^{-1}_{1}$ (green), 
$\alpha^{-1}_{2L}$ (blue), $\alpha^{-1}_{3}$ (red) and $\alpha^{-1}_{\text{B-L}}$ (black) for extra matter $\mathbf{16}_{M}+\mathbf{16}_{M}$. 
For $\Lambda_{\text{SUSY}}=1$ TeV one obtains $\Lambda_{\text{R}}\simeq 10^{15.5}$ GeV and  $\Lambda_{10}\simeq 10^{16.7}$ GeV.
The RG running of $SU(2)_{R}$ gauge coupling constant between $\Lambda_{R}$ and $\Lambda_{10}$ coincides with that of $SU(2)_{L}$. }
\label{so10}
\end{figure}

Take a pair of  $\mathbf{16}_{M}$ for example,
they contribute to $\delta b_{i}(\mathbf{M})=(4,4,4)$ and $\delta b_{i}(\mathbf{M})=(4,4,4, 4)$ in Eq.(\ref{b21}) and Eq.(\ref{b22}), respectively,
which gives rise to $\Lambda_{\text{R}}\simeq 10^{15.5}$ GeV, unified gauge coupling $\alpha^{-1}\simeq 5.45$, 
and $\Lambda_{10}\simeq 10^{16.7}$ GeV.
Fig.\ref{so10} shows the RG running of SM gauge coupling constant,
which offers us perturbative $SO(10)$ unification.
Note, the RG running of $SU(2)_{R}$ gauge coupling constant between $\Lambda_{R}$ and $\Lambda_{10}$ coincides with that of $SU(2)_{L}$,
and as required $\alpha^{-1}_{Y}$ is equal to $\frac{3}{5}\alpha^{-1}_{2R}+\frac{2}{5}\alpha^{-1}_{B-L}$ at RG scale $\Lambda_{R}$.

Repeat the analysis for other choices on extra matters in Eq.(\ref{s2}).
We find that in these cases $b$ coefficients such as 
$\delta b_{i}(\mathbf{10}_{M}+\mathbf{54}_{M})=(11.5, 11.5,11.5,11.5)$ are always 
too large to support the idea of perturbative unification.\\

\section{Conclusion}
In the forthcoming years we will enter into a new era of precise Higgs physics,
which means that studying new physics through the Higgs portal will become very interesting.
In this paper, we have utilized perturbative unification due to renormalization to explore the low energy effective theory of SUSY $SO(10)$ scenarios.
With the rational theoretical constraints taken into account,
we find that for the breaking pattern of either $\rm{SU}(5)$ or Pati-Salam 
only $\mathbf{16}$ supermultiplet can appear at TeV scale apart from the MSSM spectrum.

The quarks or leptons in the $\mathbf{16}$ supermultiplet(s) can be either chiral or vector-like. 
Note, vector-like fermion mass requires addition of SM singlet (with vev of order TeV) which does not affect our discussions. 
While the chiral case has been excluded, the vector-like quarks or leptons are smoking guns in these SUSY $SO(10)$ scenarios.
Moreover, the neutral fermions of singlet or doublets of the $\mathbf{16}$ supermultiplet can serve as dark matter totally, 
or partially with the neutralinos of the MSSM.

\begin{acknowledgments}
The author is grateful to Dr. J. Zhang for discussions. 
This work is supported in part by the National Natural Science Foundation of China under Grant No.11775039 and 
the Fundamental Research Funds for the Central Universities with project No. cqu2017hbrc1B05 at Chongqing University.
\end{acknowledgments}

\end{document}